\pdfoutput=1
\documentclass[twocolumn]{aastex631}

\usepackage{amsmath}
\usepackage{graphicx}
\usepackage{url}
\usepackage{tabularx}
\usepackage{footmisc}
\usepackage{verbatim}
\usepackage{appendix}
\usepackage{enumerate}

\shorttitle{Rescuing Kepler TCEs}
\shortauthors{Bryson et al.}

\begin{document}

\title{Rescuing Unrecognized Exoplanet Candidates in Kepler Data}

\correspondingauthor{Steve Bryson}
\email{steve.bryson@nasa.gov}

\author[0000-0003-0081-1797]{Steve Bryson}
\affiliation{NASA Ames Research Center, Moffett Field, CA 94035, USA}

\author{Kylar Flynn}
\affiliation{Marin Academy, San Rafael, CA 94901, USA}

\author{Halle Hanna}
\affiliation{Marin Academy, San Rafael, CA 94901, USA}

\author{Talia Green}
\affiliation{Marin Academy, San Rafael, CA 94901, USA}

\author[0000-0003-1634-9672]{Jeffrey L. Coughlin}
\affiliation{SETI Institute, 189 Bernardo Ave, Suite 200, Mountain View, CA 94043, USA}

\author[0000-0001-9269-8060]{Michelle Kunimoto}
\affiliation{Kavli Institute for Astrophysics and Space Research, Massachusetts Institute of Technology, Cambridge, MA 02139}

\begin{abstract}
The prime Kepler mission detected 34,032 transit-like signals, out of which 8,054 were identified as likely due to astrophysical planet transits or eclipsing binaries.  We manually examined 306 of the remaining 25,978 detections,  and found six plausible transiting or eclipsing objects, five of which are plausible planet candidates (PCs), and one stellar companion.  One of our new PCs is a possible new second planet in the KOI 4302 system.  Another new PC is a possible new planet around the KOI 4246, and when combined with a different possible planet rescued by the False Positive Working Group, we find that KOI 4246 may be a previously unrecognized three-planet system. 
\end{abstract}



\section{Introduction} \label{section:introduction}
The Kepler space telescope \citep{Borucki2010,Koch2010} has delivered unique data that enables the characterization of exoplanet population statistics, from hot Jupiters in short-period orbits to terrestrial-size rocky planets in orbits with periods up to $\sim$one year\footnote{\url{https://exoplanetarchive.ipac.caltech.edu/docs/occurrence_rate_papers.html}}. 
By observing $>$150,000 stars nearly continuously for four years looking for transiting exoplanets, Kepler detected over 4,000 planet candidates (PCs), leading to the confirmation or statistical validation of over 2,300 exoplanets.  This rich trove of exoplanet data has delivered many insights into exoplanet structure and formation, and promises deeper insights with further analysis. 

The final Kepler catalog, DR25  \citep{Thompson2018}, starts with 34,032 {\it threshold crossing events} (TCEs) \citep{Twicken2016}, indicating the detection of periodic transit-like events. 
Identification of the PCs from the TCEs was performed by a fully automated {\it Robovetter} \citep{Coughlin2016,Thompson2018}.  The Robovetter computes a variety of metrics for each TCE that measure the extent to which the TCE resembles a planetary transit or stellar eclipse.  These metrics are compared to thresholds, many of which were tuned on synthetic test datasets described below.  When a TCE passes tests that indicate a planetary transit or eclipsing binary, it is elevated to a Kepler Object of Interest (KOI). If the KOI passes further tests, it is elevated to PC status.  Such automated detection and planet candidate vetting is critical for the production of a statistically uniform catalog that is amenable to statistical correction for completeness and reliability.   The DR25 planet candidate catalog contains 4034 identified PCs out of 8054 KOIs.  


As described in \citet{Thompson2018}, the Robovetter disposition score is a measure of the confidence of the Robovetter's classification of a TCE into a PC or false positive (FP).  This score is measured by varying Robovetter metrics according to their uncertainties, and the score of a TCE is the fraction of those variations for which the TCE is classified as a PC.  
A high-score TCE (near 1.0) is almost always classified as a PC, while a low-score TCE (near 0.0) is almost always classified as an FP. 

For the DR25 catalog, a TCE is classified as a KOI if a) its Robovetter Not Transit Like (NTL) flag == 0, indicating that the transit signal is consistent with an astrophysical transit or eclipse, b) its NTL flag = 1 and its score is greater than 0.1, indicating that for at least 10\% of the uncertainty distributions of the Robovetter metrics this TCE was classified as a planet candidate, or c) it was a KOI in previous catalogs.  The result is 8054 KOIs in the DR25 KOI table, of which 4,020 are FPs and 4,034 are PCs.  Most FP KOIs are stellar eclipsing binaries or transit signals on non-target stars, with a small number of non-astrophysical false alarms.

The remaining 25,978 TCEs that did not become KOIs are mostly non-transit-like instrumental false alarms.  But the thresholds used by the DR25 Robovetter were tuned to balance completeness and reliability in a statistical sense.  Individual TCE dispositions may not be accurate because of subtleties of that TCE's signal that are not well captured by the Robovetter.  This observation motivated the Kepler False Positive Working Group (FPWG), which examined all false positives in the DR25 KOI table, resulting in the Certified False Positive Table \footnote{\url{https://exoplanetarchive.ipac.caltech.edu/cgi-bin/TblView/nph-tblView?app=ExoTbls&config=fpwg}\label{footnote:CFPTable}}, and found that many false positives are plausibly planet candidates \citep{Bryson2015}.  

In this paper we present an investigation of a subset of the 25,978 non-KOI TCEs, seeking plausible but previously unidentified planet candidates.  We do not examine DR25 KOIs because all KOIs are either planet candidates or have been manually examined by the FPWG.  In \S\ref{section:populations} we describe the TCE subsets we examine in this paper.  \S\ref{section:vetting} describes our manual vetting method and its validation.  \S\ref{section:newPCs} presents our list of plausible planet candidates.

\section{Borderline Non-KOI TCE Populations} \label{section:populations}
\begin{figure}[h!]
  \centering
  \includegraphics[width=0.98\linewidth]{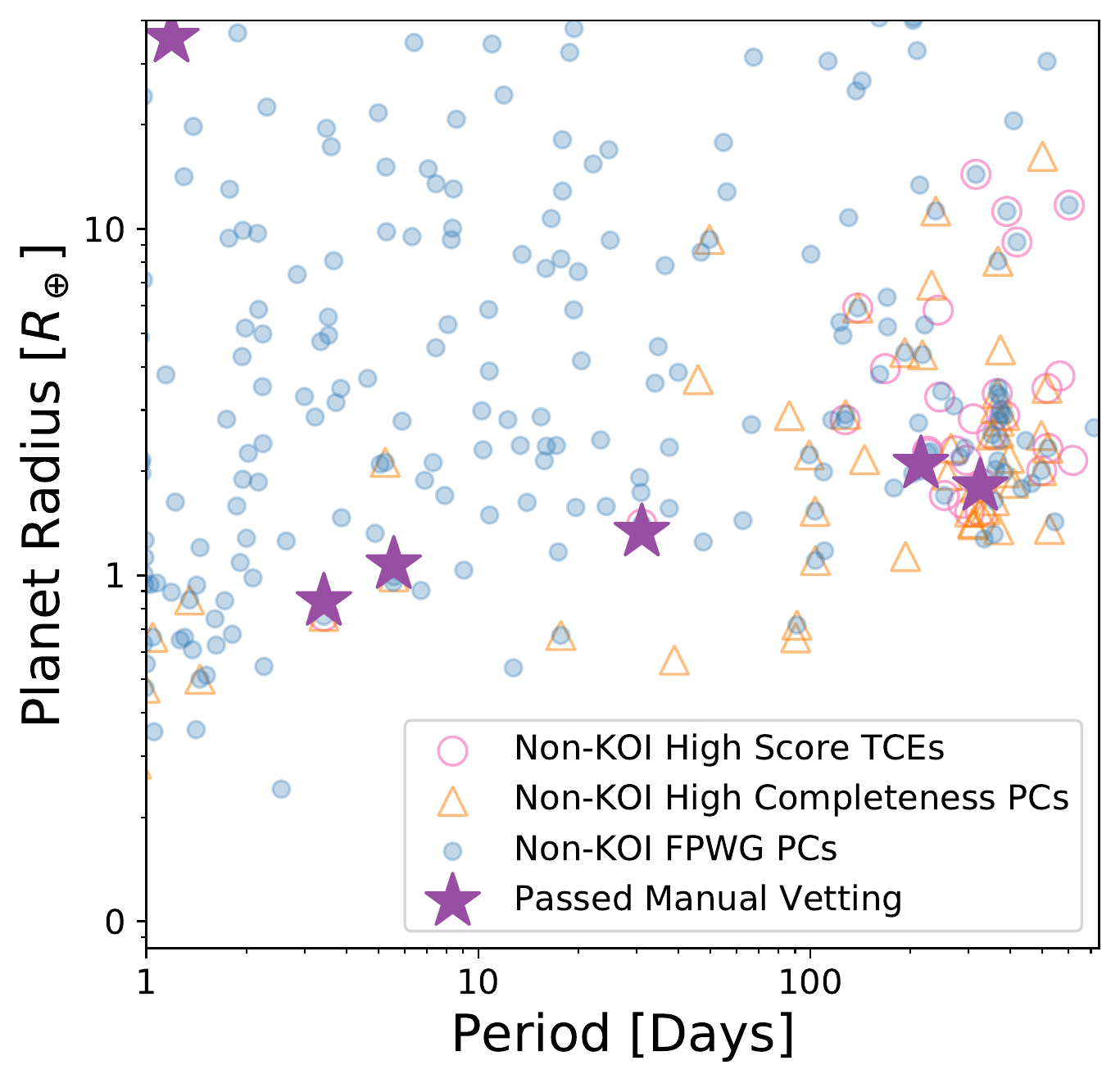} 
\caption{The TCE populations examined in this paper, with the newly recognized plausible planet candidates that passed our manual vetting.  Planet radii are based on the DR25 KIC Stellar characteristics.} \label{figure:newNonKoiPcs}
\end{figure}

Because it is not feasible to manually examine all 25,978 Non-KOI DR25 TCEs, we restricted our investigation to populations that we consider likely to have unrecognized planet candidates.  These borderline populations are selected based on either their Robovetter score or on experiments changing Robovetter selection thresholds to include more TCEs.  There is overlap between these populations, totaling 306 unique TCEs that we examined, of which six passed manual vetting and are presented in \S\ref{section:newPCs}.  The period and radii of these populations are shown in Figure~\ref{figure:newNonKoiPcs}.

\subsection{Low-Score TCEs} \label{section:lowScorePopulation}
Our first population is the set of TCEs with a Robovetter score between 0.05 and 0.1.  These TCEs were, in some sense, close to being classified as KOIs.  There are 31 non-KOI TCEs within this score range.  Three of these TCEs passed the manual vetting described in \S\ref{section:vetting}, and are described in \S\ref{section:newPCs}.

\subsection{Robovetter Variations} \label{section:robovetterVariations}

\citet{Bryson2020b} described an experiment where thresholds in the DR25 Robovetter were varied to explore alternative plausible DR25 catalogs, to study the effect of completeness and reliability corrections.  Here we apply two sets of Robevetter thresholds from Bryson et al. (2020)
to the non-KOI TCEs.  This produced several new Robovetter PCs from the non-KOI TCEs:
\begin{itemize}
    \item {\bf High Completeness Thresholds \footnote{\url{https://github.com/stevepur/DR25-occurrence-public/blob/main/GKRobovetterVariations/koiCatalogs/dr25_GK\_KOIs\_highCompleteness.csv}}:} Relax the DR25 Robovetter thresholds to allow more planet candidates, which will also allow more false positives and therefore have lower reliability.  58 of the PCs in this catalog are not DR25 KOIs
    \item {\bf FPWG PC Thresholds\footnote{\url{https://github.com/stevepur/DR25-occurrence-public/blob/main/GKRobovetterVariations/koiCatalogs/dr25\_GK\_KOIs\_fpwgpc.csv}}:} Tune the DR25 Robovetter thresholds to allow those DR25 KOIs identified as false positive, but were marked as possible planets in the Certified False Positive table on the exoplanet archive, to now be labeled as PC by the Robovetter.  This resulted in a catalog with higher completeness and lower reliability, but with different planet candidates compared with the the high completeness thresholds \citep[see][]{Bryson2020b}. 270 of the PCs in this catalog are not DR25 KOIs
\end{itemize}
In \citet{Bryson2020b}, these catalogs were restricted to DR25 Kepler KOIs and contained 97 planet candidates that had been dispositioned as FP by the Robovetter using DR25 thresholds.  \citet{Bryson2020b} did not examine these new PCs individually, but treated them statistically as a population.  Because they were DR25 KOIs, they were manually examined by the FPWG, and of these 97 new planet candidates, 37 were classified as possible planets, 30 were certified as false alarms, 16 certified as false positives, and 14 were  inconclusive\footref{footnote:CFPTable}.  So most of the new planet candidates in the high-completeness and FPWG PC catalogs are not true planets.  Of the 37 new PCs, 24 have DR25 Robovetter score = 0, indicating that Robovetter score is not a reliable indication that any given individual TCE is a planet.  

There are 58 non-KOI Robovetter PCs in the high-completeness catalog and 270 non-KOI Robovetter PCs in the FPWG PC catalog.  Three of these TCEs passed the manual vetting described in \S\ref{section:vetting}, and are plausible planet candidates as described in \S\ref{section:newPCs}.  We believe that the rest of the non-KOI PCs in these alternative catalogs are unlikely to be true planets.

\section{The Manual Vetting Process} \label{section:vetting}
The TCEs described in \S\ref{section:populations} were subjected to a manual vetting process, using a simplified version of the method used to certify false positives in the Kepler certified false positive table \citep{Bryson2015}, described in \S\ref{section:vettingProcess}.  As described in \S\ref{section:vettingValidation} we applied the same vetting method to a set of known PCs and FPs, with similar properties to validate our vetting method.

\subsection{Vetting Process} \label{section:vettingProcess}
Our vetting process starts with the visual examination of the q1\_q17\_dr25\_tce report for each TCE, found on the exoplanet archive.  This report presents the TCE's light curve from a variety of perspectives, as well as difference-image pixel analysis indicating whether the TCE detection occurrs on the target star.  TCEs were triaged by asking two questions: is the detected event transit-like, and is it on the target star?

The transit-like nature of the TCE was determined by examining folded light curves for all Kepler observations, as well as folded light curves within each quarter.  Because these TCEs are mostly low SNR, we do not expect the transits to always be obvious in individual quarters.  A TCE can be considered transit-like even if it cannot be seen in individual quarters, so long as the transit signal becomes more obvious when the quarters are combined and folded.  However, many false alarm detections are due to single deep events that align with fluctuations to trigger a TCE detection.  Therefore a TCE was considered not transit-like at the TCE's ephemeris if there is a deep transit-like event in one quarter but no similarly deep events in other quarters.  We also required that the transit be unique in the sense that there is no other transit-like event with the same period but different phase, based on the model-shift uniqueness test in the TCERT report.  
For more details on these transit-like requirements, see \S A.3 of \citet{Thompson2018}.

The determination that the TCE is on the target star used the difference image technique described in \citet{Bryson2013}.  For these low-SNR TCEs, it was often the case that the PRF-fitted centroid component of this technique performed poorly, in which case we rely on visual examination of the difference images in the TCERT report.  Consistent with Kepler vetting, we use an ``innocent until proven guilty'' approach: when the difference image indicates that the pixels with the largest change correspond to the target star, the TCE is considered to be on the target star.  This does not guarantee that the TCE event is on the target star due to the possibility of unresolved field stars.

Based on the examination of the TCE reports, the TCEs were assigned a grade of bad (obvious false positive or alarm), poor (likely false positive or alarm), medium (the data are ambiguous) and good (transit-like and on the target star).  Some TCEs were given mixed grades such as medium good.  

TCEs with a grade of good or medium-good were subjected to further examination of their folded light curves using \texttt{Lightkurve} software \citep{lightkurve2018}.  Both raw simple aperture photometry and light curves corrected by \texttt{Lightkurve} using co-trending basis vectors, which remove common signals, were examined.  One purpose of this examination is to examine how the folded light curve appears in phase windows that are not available in the TCERT reports, further confirming the uniqueness of the transit signal.  Another purpose is to verify that the transit signal is not sensitively dependent on the level of binning used to average the folded data.  These low SNR TCEs often require significant binning to make the transit signal visually apparent.  When the TCE passes these tests, we consider it to be a plausible planet candidate.

\subsection{Validation of the Vetting Process} \label{section:vettingValidation}
To test and validate our manual vetting process, we performed the vetting described in \S\ref{section:vetting} on data sets with known populations of planets and false positives.  For data with known planets we used TCES from the injected data set \citep{Christiansen2020}.  For populations with no planet candidates, so all TCEs are false positives, we use the inverted and scrambled data sets (see \S2.3 of \citet{Thompson2018} and \citet{Coughlin2017}).  There was no selection based on the strength of the transit signal in the injected planet data, so we expect that some of the injected planets will not be high enough S/N to be recognizable as transits.  Because the injected planets are all on the target star, this validation method only tests the transit-like judgement of our vetting process.  

Because it is not possible to perform light curve analysis without knowing the source of the TCE's data, breaking the double-blind nature of our validation, we limited our validation to the triage phase.  All the observed TCEs that were graded good passed light curve validation.  

We performed our validation in a double-blind form.  One team member, who did not take part in the triage phase described in \S\ref{section:vetting}, wrote a program that randomly selects TCEs from the injected, inverted, and scrambled data sets, then assembles their TCE reports and scrubs from the TCE reports information identifying the source of the TCE.  Such scrubbing was only possible when a TCE was the only TCE detected on its host star, so systems with multiple TCEs were not used in this test.  31 TCEs were selected in this way, which were vetted by the other team members using the same triage criteria applied to observed data.  

Seven simulated-data planets were graded as good, of which five were correct because they were from the injected planet data and two were incorrect because they were from the inverted or scrambled data. One TCE from the inverted data set that was incorrectly graded good turned out to be from a flare star, so under inversion the flare signals appeared as plausible transit signals, explaining the incorrect identification as a PC.  The other incorrectly identified ``good'' PC, from the scrambled data set, was judged to be sufficiently convincing that most human vetters would have made the same error.  

The incorrectly vetted PC signals were low SNR, long-period signals that are in the low-reliability regime of the DR25 Robovetter \citep{Thompson2018}.  We conclude that our vetting method, identifying 5 out of 6 for 83\% accuracy is comparable to human vetting and the DR25 Robovetter for low SNR signals.  Because we are claiming new plausible PCs, we are not concerned with the completeness of our vetting method.

\section{New Planet Candidates} \label{section:newPCs}
\begin{figure*}[ht]
  \centering
  \includegraphics[width=0.92\linewidth]{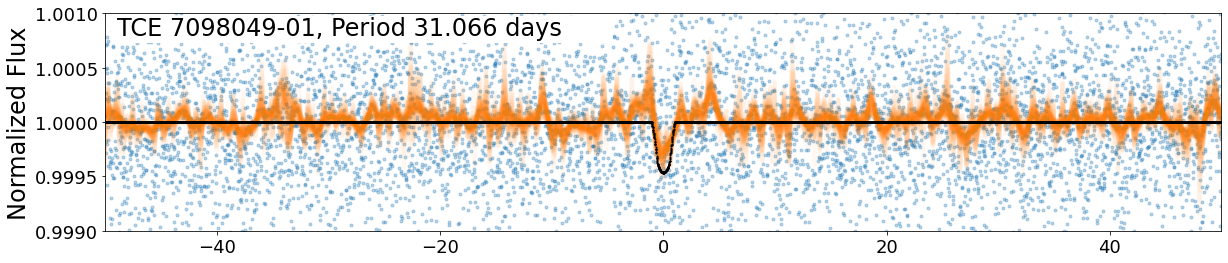} 
  \includegraphics[width=0.92\linewidth]{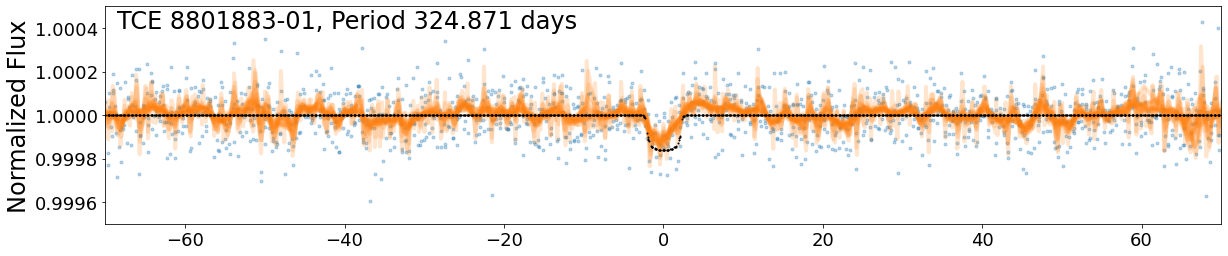} 
  \includegraphics[width=0.92\linewidth]{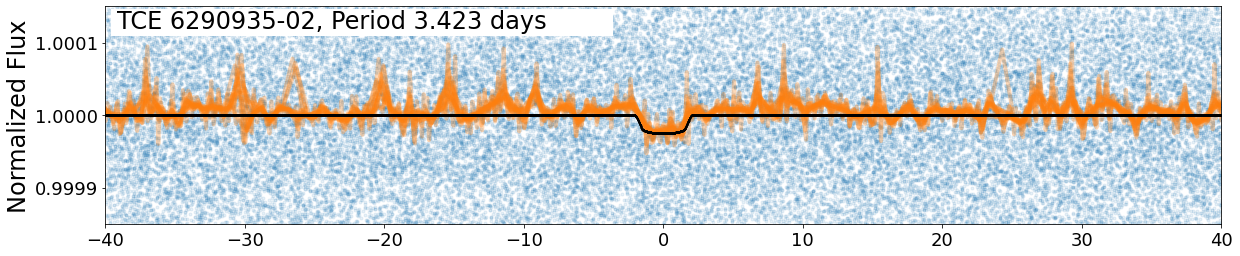}
  \includegraphics[width=0.92\linewidth]{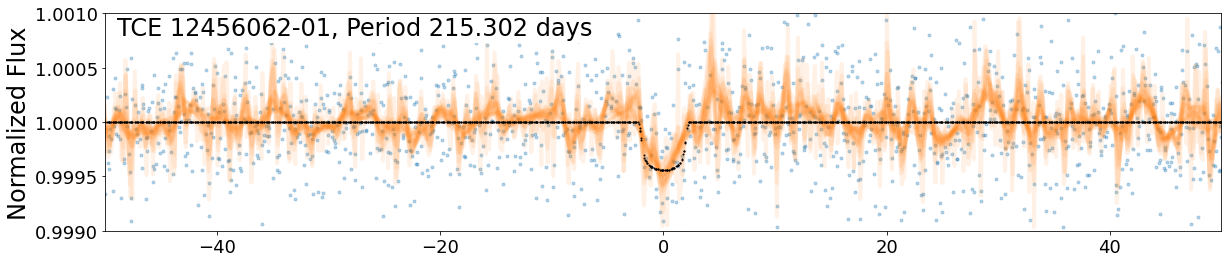}
  \includegraphics[width=0.92\linewidth]{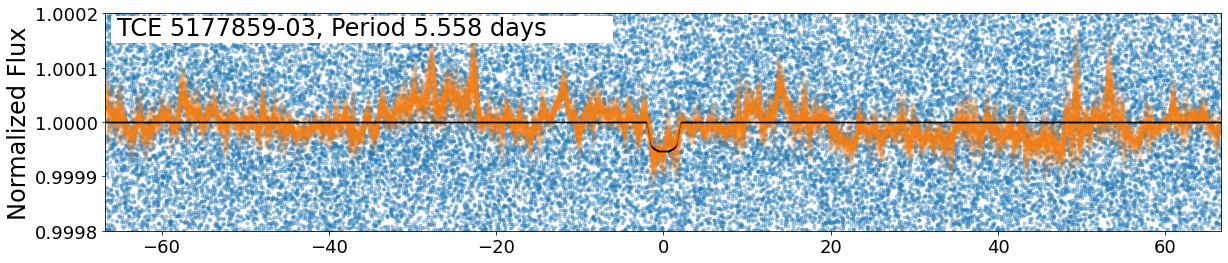}
  \includegraphics[width=0.92\linewidth]{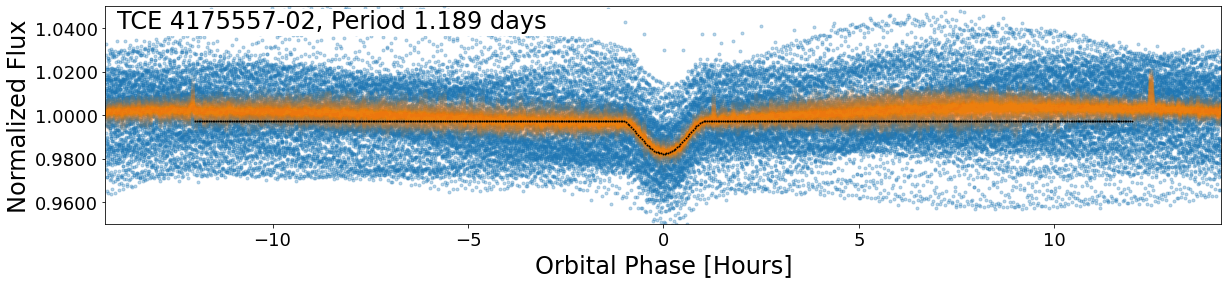}
\caption{Lightcurves for the newly recognized planet candidates and eclipsing binary (TCE 004175557). Blue points are the folded detrended photometric data; Orange points are binned photometric data for a variety of binnings, and the black line is the transit model.  These figures are among several diagnostics used to identify new planet candidates. See text for details.} \label{figure:lightcurves}
\end{figure*}

We identify five new plausible PCs and one new short-period binary star from the populations described in \S\ref{section:populations}, summarized in Table~\ref{table:PCs}.  We do not claim confidence that the new PCs are actually planets, just that they are as likely to be planets as DR25 planet candidates with similar period and radius.  For each new PC we provide the planet properties from the DR25 TCE table and analysis, with the exception of TCE 004175557-02, which is treated as a special case in \S\ref{section:004175557}.  We also provide updated planet radii based on the Gaia-derived stellar properties from \citep{Berger2020a}.  The light curves for these PCs are shown in Figure~\ref{figure:lightcurves}, generated using the \texttt{Lightkurve} package.  To mitigate binning artefacts, we show a distribution of binnings for each light curve.  The transit model from the DR25 analysis is shown for each TCE, with the exception of TCE 004175557-02 where we fit our own transit model as described in \S\ref{section:004175557}.  In this section we compare the measured transit duration with that expected from the formulae in \citet{Seager2003} as a qualitative diagnostic of the consistency of the fitted planet properties.

The first three new PCs, TCE 007098049-01, TCE 008801883-01, and TCE 006290935-02, are from the low-score population of \S\ref{section:lowScorePopulation}.  The other three, TCE 012456062-01, TCE 005177859-03, and TCE 004175557-02 are from the FPWG PC population of \S\ref{section:robovetterVariations}.  We did not find any plausible PCs in the high-completeness population of \S\ref{section:robovetterVariations}. 

\begin{table*}[ht]
\movetableright=-0.35in
\caption{Our new planet candidates, with radius based on Gaia stellar parameters}\label{table:PCs}
\begin{tabular}{r c c c c c c}
\hline
\hline
TCE ID & Period (Days) & Depth (ppm) & Duration (hours) & b & Radius ($R_\oplus$) & RV Score \\
\hline
007098049-01 & 31.06558 & 464.5 & $1.62 \pm 0.34$ & $0.20 \pm 34.24$ & $1.11 \pm 2.57$ & 0.090 \\
008801883-01 & 324.8712467 & 160.7 & $4.76 \pm 0.94$ & $0.93 \pm 0.15$ & $2.73 \pm 0.51$ & 0.086 \\
006290935-02 & 3.422508 & 25.2 & $3.64 \pm 0.51$ & $0.89 \pm 0.63$ & $0.87 \pm 0.44$ & 0.051 \\
012456062-01 & 215.301708 & 441.8 & $4.14 \pm 2.75$ & $0.65 \pm 9.66$ & $2.19 \pm 4.78$ & 0.007 \\
005177859-03 & 5.558420 & 53.9 & $3.78 \pm 0.5442$ & $0.90 \pm 0.40$ & $0.76 \pm 0.32$ & 0.090 \\
\end{tabular}
\end{table*}

\subsection[TCE 007098049-01]{TCE 007098049-01\footnote{\url{https://exoplanetarchive.ipac.caltech.edu/data/KeplerData/007/007098/007098049/tcert/kplr007098049_q1_q17_dr25_obs_tcert.pdf}}} \label{section:007098049}

TCE 007098049-01 was detected at a period of 31.065582 days, with a depth of 464.5 ppm, a duration of $1.62 \pm 0.34$ hours, and a DR25 Robovetter score of 0.090. KIC 007098049, the host star, has an effective temperature of
$4511^{+121}_{-135}~K$, radius $0.64^{+0.05}_{-0.06}~R_\odot$, and mass $0.62^{+0.07}_{-0.04}~M_\odot$ in the Kepler Input Catalog, based on photometric characterization.  Using these KIC stellar parameters, this TCE implies a planet with radius $1.33 \pm 3.08 ~ R_\oplus$ and a very uncertain impact parameter of $b=0.20 \pm 34.24$.  The observed transit duration is consistent with the predicted transit duration of 1.49 hours assuming $b=0.20$.  The Gaia-based radius of this star from \citep{Berger2020a} is $0.532^{+0.015}_{-0.014}~R_\odot$ and mass is $0.556^{+0.018}_{-0.018}~M_\odot$, implying a planet radius of $1.11 \pm 2.57 ~ R_\oplus$, assuming no change in the transit fit.

The transit signal is weak in individual DV Quarter-Phased Transit Curves, consistent with a low SNR of $7.6$. Stronger and better-defined transit signals than those in individual quarters are present in both the seasonal sums, particularly S2, and in the yearly sums. The summed transit curve for all quarters also reveals a well-defined transit signal.

The transit signal also proved robust against changes in binning. Seen in Fig. 1, the light curve for TCE 007098049-01 shows minimal variation for binning factors ranging from 10 to 80.

\subsection[TCE 008801883-01]{TCE 008801883-01\footnote{\url{https://exoplanetarchive.ipac.caltech.edu/data/KeplerData/008/008801/008801883/tcert/kplr008801883_q1_q17_dr25_obs_tcert.pdf}}} \label{section:008801883}

TCE 008801883-01 was detected at a period of 324.8712467 days, with a depth of 160.7 ppm, and a duration of $4.76 \pm 0.94$ hours and a DR25 Robovetter score of 0.086. Its host star, KIC 8801883, has an effective temperature of $6788^{+161}_{-242}~K$, radius $1.19^{+0.25}_{-0.14}~R_\odot$, and a mass of $1.16^{+0.13}_{-0.16}~M_\odot$, based on photometric characterization supplemented by $\log(g)$ from the flicker technique \citep{Bastien2016}.  Using these KIC stellar parameters, this TCE implies a planet with radius $1.81 \pm 0.50 ~ R_\oplus$ and impact parameter $b=0.93 \pm 0.15$.  The Gaia-based radius of this star from \citep{Berger2020a} is $1.80^{+0.05}_{-0.05}~R_\odot$ and mass is $1.33^{+0.06}_{-0.05}~M_\odot$, implying a planet radius of $2.73 \pm 0.51 ~ R_\oplus$, assuming no change in the transit fit.

The transit signals are relatively well defined and consistent in individual quarters. The seasonal and yearly sums produce strong, summed transit signals. 

The predicted duration of 3.76 hours given b=0.93 using the KIC stellar parameters is about one sigma from the observed duration.  An impact parameter of about $b=0.87$ yields the observed transit duration.

\subsection[TCE 006290935-02]{TCE 006290935-02\footnote{\url{https://exoplanetarchive.ipac.caltech.edu/data/KeplerData/006/006290/006290935/tcert/kplr006290935_q1_q17_dr25_obs_tcert.pdf}}} \label{section:006290935}
TCE 006290935-02 was detected with a period of 3.422508 days, a depth of 25.2 ppm, and a duration of $3.64 \pm 0.51$ hours and a DR25 Robovetter score of 0.051. The host star KIC 6290935 has an effective temperature of $6170\pm83~K$, radius $1.43^{+0.25}_{-0.28}~R_\odot$ and mass $1.05^{+0.09}_{-0.08}~M_\odot$.   Using these KIC stellar parameters, this TCE implies a planet with radius $0.84 \pm 0.45 ~ R_\oplus$ and impact parameter $b=0.89 \pm 0.63$.  The Gaia-based radius of this star from \citep{Berger2020a} is $1.47^{+0.04}_{-0.04}~R_\odot$ and mass is $1.13^{+0.07}_{-0.08}~M_\odot$, implying a planet radius of $0.87 \pm 0.44 ~ R_\oplus$, assuming no change in the transit fit.

The transit signals in the individual DV Quarter-Phased Transit Curves largely appear weak and inconsistent, as expected for an SNR of 7.8. Q14 shows a stronger transit-like signal than the other quarters, but yearly and seasonal sums that do not include Q14 show improvement over the individual quarters, with increasingly prominent transit signals as more quarters are summed. The summed transit curve for all quarters is well-defined.  The transit signal survived further examination using \texttt{Lightkurve}, proving robust across a range of binning factors from 100 to 800, shown in Fig. 3. 

The expected transit duration of 2.70 hours given the KIC stellar parameters and fitted impact parameter $b=0.89$ is more than one sigma from the observed duration.  An impact parameter of about $b=0.80$ predicts about 3.62 hours, close to the observed transit duration.

TCE 006290935-01 = KOI 4302.01 is a DR25 planet candidate with radius $1.24 \pm 0.62 ~ R_\oplus$ ($1.28 \pm 0.59 ~ R_\oplus$ using \citet{Berger2020a} stellar parameters) and period 5.53 days.  




\subsection[TCE 012456062-01]{TCE 012456062-01\footnote{\url{https://exoplanetarchive.ipac.caltech.edu/data/KeplerData/012/012456/012456062/tcert/kplr012456062_q1_q17_dr25_obs_tcert.pdf}}} \label{section:012456062}

TCE 012456062-01 was detected at a period of 215.301708 days, with a depth of 441.8 ppm, and a duration of $4.14 \pm 2.75$ hours and a DR25 Robovetter score of 0.007. KIC 012456062, the host star, has an effective temperature of
$6092^{+163}_{-200}~K$, radius $0.94^{+0.30}_{-0.10}~R_\odot$, and mass $1.02^{+0.14}_{-0.14}~M_\odot$ in the Kepler Input Catalog, based on photometric characterization.  Using these KIC stellar parameters, this TCE implies a planet with radius $2.09 \pm 4.62 ~ R_\oplus$ and highly uncertain impact parameter $b=0.65 \pm 9.66$.  The predicted transit duration using $b=0.65$ is 3.28 hours, consistent with the observed duration, while $b=0.41$ predicts 4.19 hours.  The Gaia-based radius of this star from \citep{Berger2020a} is $0.98^{+0.03}_{-0.03}~R_\odot$ and mass is $1.00^{+0.47}_{-0.06}~M_\odot$, implying a planet radius of $2.19 \pm 4.78 ~ R_\oplus$, assuming no change in the transit fit.

\subsection[TCE 005177859-03]{TCE 005177859-03\footnote{\url{https://exoplanetarchive.ipac.caltech.edu/data/KeplerData/005/005177/005177859/tcert/kplr005177859_q1_q17_dr25_obs_tcert.pdf}}} \label{section:005177859}

TCE 005177859-03 was detected at a period of 5.558420 days, with a depth of 53.9 ppm, and a duration of $3.78 \pm 0.5442$ hours and a DR25 Robovetter score of 0.090. KIC 005177859, the host star, has an effective temperature of
$5839^{+156}_{-174}~K$, radius $1.227^{+0.360}_{-0.270}~R_\odot$, and mass $1.044^{+0.137}_{-0.125}~M_\odot$ in the Kepler Input Catalog, based on photometric characterization.  Using these KIC stellar parameters, this TCE implies a planet with radius $1.073 \pm 0.53 ~ R_\oplus$ and impact parameter $b=0.90 \pm 0.40$.  The Gaia-based radius of this star from \citep{Berger2020a} is $0.9356^{+0.017}_{-0.016}~R_\odot$ and mass is $1.0113^{+0.024}_{-0.031}~M_\odot$, implying a planet radius of $0.76 \pm 0.32 ~ R_\oplus$, assuming no change in the transit fit.

The expected transit duration of 2.84 hours given the KIC stellar parameters and fitted impact parameter $b=0.90$ is more than one sigma from the observed duration.  However, an impact parameter of about $b=0.80$ predicts about 3.85 hours, close to the observed transit duration.

TCE 005177859-02 = KOI 4246.02 is a DR25 planet candidate with period 8.756238 days and radius $1.36 \pm 0.70  ~ R_\oplus$, and TCE 005177859-01 = KOI 4246.01 is a DR25 false positive that was judged to be a possible planet candidate by the FPWG with period 6.984718 days and radius $1.52 \pm 0.48  ~ R_\oplus$. 
So KOI 4246 may be a previously unrecognized three-transiting-planet system.  The occurrence of three plausible planet candidate signals on KOI 4246 increases our confidence that TCE 005177859-03 is a planet candidate via the multiplicity boost of \citet{Rowe2014}.

\subsection[TCE 004175557-02]{TCE 004175557-02\footnote{\url{https://exoplanetarchive.ipac.caltech.edu/data/KeplerData/004/004175/004175557/tcert/kplr004175557_q1_q17_dr25_obs_tcert.pdf}\label{footnote:004175557}}} \label{section:004175557}

TCE 004175557-02 was detected at a period of 1.18948 days, with a depth of 461.4 ppm, and a duration of $1.37\pm0.49$ hours and a DR25 Robovetter score of 0.090. TCE 004175557-01 is an obvious false alarm due to stellar variability.  The host star KIC 004175557 has an effective temperature of $5680^{+169}_{-186}~K$, radius $0.91^{+0.25}_{-0.09}~R_\odot$, and mass $0.96^{+0.10}_{-0.10}~M_\odot$ in the Kepler Input Catalog, based on photometric characterization.  However the planet properties from the Kepler pipeline ($R_p = 2.35 \pm 1.13 ~ R_\oplus$) are very likely wrong. The host star exhibits an $\approx 5.4\%$ sinusoidal variation with a measured period of $1.185$ days \citep{McQuillan2014}.  Though this sinusoidal variation's period is close to the transit signal period, it is not exactly the same, and the phase of the transit signal changes significantly relative to the phase of the sinusoid over the Kepler observations.  As is apparent in the TCE report\footref{footnote:004175557}, the Kepler pipeline did a poor job of removing the sinusoidal signal, leaving an oscillatory, not-transit-like signal. The non-detrended data, however, show a clean, v-shaped transit signal with period 1.18948 days, and the alternative detrending analysis \citep{Li2019} in the TCE report cleanly brings out that signal with a depth of 16,000 ppm.  A phase curve and small (600 ppm) secondary are also apparent in the alternative detrended light curve.

The TCE report presents a fit to the alternative detrended light curve, resulting in a planet radius $13.17 \pm 2.0 ~ R_\oplus$. The Gaia-based radius of this star from \citep{Berger2020a} is $1.23^{+0.063}_{-0.059}~R_\odot$ and mass is $0.9231^{+0.07}_{-0.06}~M_\odot$, implying a planet radius of $17.77 \pm 2.7 ~ R_\oplus$, assuming no change in the transit fit.  Based on the analysis of \citet{Coughlin2012}, a planet with this size and an albedo of $\approx 0.8$ explains the secondary.  However, based on these planet parameters, we expect a duration of 1.77 hours, notably shorter than the duration of 2.3 hours in the fitted light curve.  This discrepancy between the light curve fit and the estimated transit duration may be due to the fact that the fit to the alternative detrended light curve uses a trapezoidal transit model \citep{Li2019}, which does not allow impact parameters greater than 1.  

Using the Batman transit modeling package \citep{Kreidberg2015}, we performed a fit of radius, inclination and semi-major axis to folded data detrended by \texttt{Lightkurve} (based on a Savitsky-Golay filter), shown in the bottom panel of \ref{figure:lightcurves}.  This fit yielded a planet radius of $35.57^{+11.07}_{-5.90} ~  R_\oplus$ ($47.99^{+15.78}_{-7.96} ~  R_\oplus$ using \citet{Berger2020a} stellar properties), inclination $67.37^{+1.16}_{-1.81}$ degrees corresponding to a median impact parameter of $b=1.19$, and a ratio of the semi-major axis to the stellar radius of $3.09^{+0.04}_{-0.02}$. Our fit used uninformative large flat priors, and was seeded with a radius of $10 ~ R_\oplus$ and ratio of the semi-major axis to the stellar radius of 4.  The fitted radius of TCE 004175557-02 is over the maximum planet size of $30 ~ R_\oplus$ used by the FPWG.  Using these planet parameters, we obtain an expected duration of 2.6 hours, more consistent with the observed duration of 2.46 hours in the fitted light curve.

We therefore conclude that TCE 004175557-02 is a previously unrecognized stellar companion.

\clearpage
\section{Discussion} \label{section:discussion}

Vetting a large collection of detections, such as the Kepler DR25 TCE table of 34,032 transit-like signals, to identify planets is impossible to do with complete accuracy.  The Kepler DR25 catalog was carefully designed to be statistically well-characterized in terms of completeness and reliability, in order to support statistical studies such as the calculation of planet occurrence rates.  This was accomplished by creating the catalog in a uniform, replicable way via an automated Robovetter that balanced completeness and reliability  \citep{Thompson2018}.   

In balancing completeness and reliability, the DR25 Robovetter inevitably misclassifies individual signals.  Of the 25,978 signals rejected as non-astrophysical false alarms by the Robovetter, we investigated 306 borderline cases and found that six, or 2\%, pass human examination as astrophysical transits or eclipses.  This indicates that, statistically speaking, the Robovetter is doing an excellent job.  However, in individual cases, the Robovetter does miss planet candidates that would have been passed by a human.

We are not in a position to estimate the impact of our new planet candidates on demographic studies, because such studies depend critically on knowing the reliability and completeness of the planet candidate catalog.  The vetting methods we used in this paper do not have such well-characterized completeness and reliability.  But two of our new planet candidates are in the small-radius, long-period regime where there are relatively few planet candidates.  An improved Robovetter that included these planets without suffering from decreased reliability may yield interestingly different, or at least more precise, demographic results.

We are also not in a position to speculate on whether an investigation of all 25,978 non-KOI TCE detections would reveal an overall 2\% rate of new planet candidates.  But we expect that there are populations beyond the 306 that we examined that may yield a similar bounty of unrecognized planet candidates.  We look forward to automated searches of the Kepler TCEs with improved sensitivity that may reveal new planet candidates.

\section{Acknowledgements} \label{section:acknowledgements}
We thank Lauren Weiss and Doug Caldwell for valuable discussions, our anonymous reviewer for helpful comments, and Stori Oates and Mary Kay Dolejsi for their support through the Marin Academy Research Collaborative.  

\software{Lightkurve \citep{lightkurve2018}, Batman \citep{Kreidberg2015}, emcee \citep{ForemanMackey2013}}

\bibliography{refs}


\end{document}